\documentclass[conference]{IEEEtran}
\IEEEoverridecommandlockouts
\usepackage{cite}
\usepackage{amsmath,amssymb,amsfonts}
\usepackage{algorithmic}
\usepackage{graphicx}
\usepackage{textcomp}
\usepackage{xcolor}
\def\BibTeX{{\rm B\kern-.05em{\sc i\kern-.025em b}\kern-.08em
    T\kern-.1667em\lower.7ex\hbox{E}\kern-.125emX}}
\begin{document}


%

\newcommand*{\Scale}[2][4]{\scalebox{#1}{$#2$}}%
\newcommand*{\Resize}[2]{\resizebox{#1}{!}{$#2$}}%
\newcommand{\BlueText}[1]{\textcolor{blue}{#1}}
\newcommand{\remove}[1]{ }

\makeatletter
\setlength{\@fptop}{0pt}
\makeatother

\newcommand{\claudio}[1]{\textcolor{blue}{{#1}}}
\newcommand{\henry}[1]{\textcolor{red}{{#1}}}

\newcommand{\R}{\mathbb{R}}
\newcommand{\Z}{\mathbb{Z}}

\title{Modeling Data Analytics Architecture for Smart Cities Data-Driven Applications using DAT}

\author{
\IEEEauthorblockN{, {Moamin B. Abughazala}\thanks{moamin.abughazala@graduate.univaq.it}\IEEEauthorrefmark{2}, {Henry Muccini}\thanks{henry.muccini@univaq.it}\IEEEauthorrefmark{2},}

\IEEEauthorblockA{\IEEEauthorrefmark{2}\textit{DISIM Department},
\textit{University of L'Aquila},
L'Aquila, Italy}
}

\maketitle

\begin{abstract}
Extracting valuable insights from vast amounts of information is a critical process that involves acquiring, storing, managing, analyzing, and visualizing data. Providing an abstract overview of data analytics applications is crucial to ensure that collected data is transformed into meaningful information. One effective way of achieving this objective is through Data Architecture. This article shares our experiences in developing a Data Analytics Architecture (DAA) using model-driven engineering for Data-Driven Smart Cities applications utilizing DAT.
\end{abstract}

\begin{IEEEkeywords} Analytics Architecture, Big Data Architecture, IoT, Smart Cities, Data-Driven
\end{IEEEkeywords}

\section{Introduction}

Analyzing data for Smart Cities requires utilizing various methods and technologies to collect, store, and examine the information produced by connected devices and sensors. This aids in comprehending the functioning and conduct of Smart Cities systems and identifying patterns and trends in the data that can enhance their efficacy \cite{ahmad2021review} \cite{simmhan2016big}. 

Smart Cities data analytics comprises three primary approaches: real-time, predictive, and prescriptive analytics \cite{elijah2018overview}  \cite{balusamy2021big}. Real-time analytics \cite{gillet2021lambda+} utilizes algorithms and software to swiftly analyze data generated by IoT devices, promptly identifying and responding to data trends and patterns. Predictive analytics, on the other hand, harnesses historical data and machine learning algorithms to anticipate future outcomes and behaviors. Lastly, prescriptive analytics employs optimization algorithms to suggest actions and decisions that can assist businesses in reaching their objectives.

Data analytics has numerous applications in the Smart Cities domain, such as predictive maintenance, energy management, supply chain optimization, and customer behavior analysis [4]. By utilizing data analytics, we can obtain valuable insights into the performance and behavior of Smart Cities systems, which can help improve their operations, reduce costs, and drive innovation.

Analytics architectures refer to the systems and technologies utilized for collecting, storing, processing, and analyzing data to obtain insights and make data-driven decisions. These architectures commonly involve using distributed computing systems, such as server clusters or cloud-based platforms, to manage data storage and processing. Along with these systems, analytics architectures also comprise specialized software tools and algorithms intended for analyzing and acquiring insights from data. These tools may comprise data mining, machine learning algorithms, and visualization and reporting tools for presenting the analysis results. An analytics architecture's specific components and design may vary depending on the organization's objectives and requirements.

This paper showcases the efficacy of modeling languages in crafting an Analytics Data Warehouse for Data-Driven Smart Cities applications, substantiated by a detailed case study.

The paper is organized as follows. The background is presented in Section II. The applied real case study is in Section III, and the conclusions are finally drawn in Section IV.

\section{Background}

\subsection{Big Data Analytics}

Big Data analytics (BDA) and Business Intelligence (BI) are two crucial fields that empower businesses to make more informed decisions. BDA involves analyzing large amounts of raw data to uncover valuable insights, such as unidentified relationships and market trends. Skilled data scientists carry out this process to provide businesses with the information they need for making strategic decisions. However, BI is a highly specialized field that uses various technologies, applications, and practices to gather, store, access, and analyze data. With BI, businesses can identify trends, and data patterns, and forecast future outcomes, enabling them to make informed decisions.

\subsection{DAT: Data Architecture Framework }
The DAT tool \cite{10.1007/978-3-031-36889-9_8}  is ideal for modeling data architecture in IoT applications \cite{abughazala2023modeling}. It clearly explains how data flows through the system and provides a blueprint for it . Stakeholders can describe two levels of data architecture: high-level architecture (HLA) and low-level architecture (LLA). It represents the data from source to destination, including formats, processing, storage, analysis, and consumption methods. The tool is built based on a structural and behavioral meta-model to support the documentation of Data-Driven applications. The data-view architecture modeling approach follows the IEEE/ISO/IEC 42010 standard \cite{42010} and uses the Data architecture structural and behavioral view (DAML) modeling language. DAT is considered the fourth view for CAPS \cite{sharaf2017architecture} \cite{sharaf2018arduino} \cite{muccini2017caps} \cite{sharaf2017simulating}.

\section{Application of DAT Models to the Analytics Data Warehouse Case Study}
\label{sec:casestudy}

This section introduces The Analytics Data Warehouse (ADW) case study and its DAT.

\begin{center} 
   \begin{figure*}[!h]
	\centering
	\makebox[\textwidth]
	{
	    	\includegraphics[width=1\paperwidth]{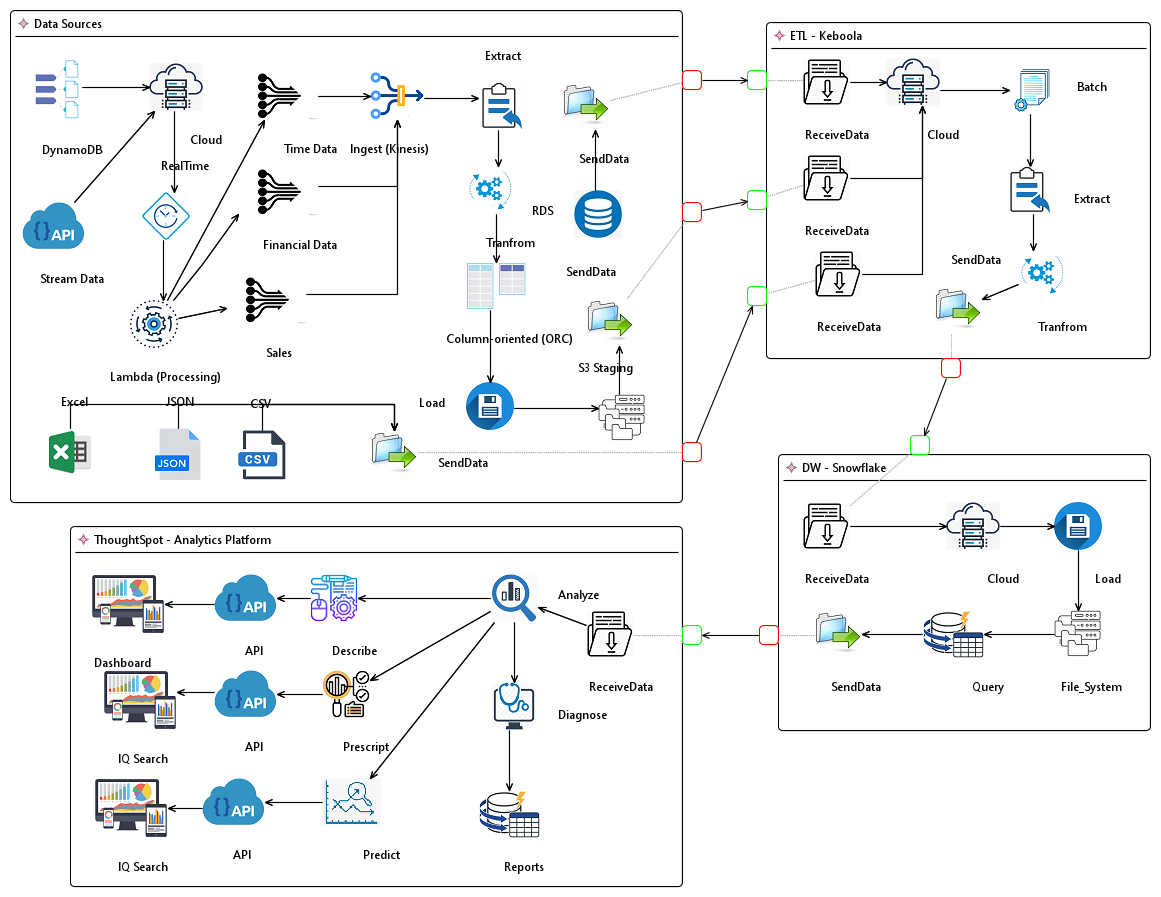}
	}
	\caption{Modeling of the Analytics Data Warehouse }
	\label{fig:ADW}
    \end{figure*}
\end{center}
\subsection{The Analytics Data Warehouse}

The data sources of this Analytics Data Warehouse are the AWS RDS (Relational Database Service) instances, the duplicate staging ORC files saved on S3 that are generated from the data streams. In addition, this data warehouse consumes data from CSV files, worksheets, and JSON objects that augment the data in the warehouse to make the data rich for analytics purposes. 
The ETL process in this data warehouse is developed and run using a cloud-based third-party tool called Keboola Connectors. This tool provides all the needed connectors for data sources like AWS MySQL RDSs, ORC(Optimized Row Columnar) file extractors, Google Drive sheets, CSV readers, etc. Keboola also offers the required environment and tools to perform data transformations on the extracted data. Moreover, it provides the needed capabilities to schedule executing all extractions and transformation and orchestration between them. In the case implementation of this data warehouse, the data prepared using Keboola is stored on the cloud-based Snowflake data warehouse. The data in the Snowflake data warehouse is consumed by another cloud-based advanced BI and analytics framework called ThoughtSpot. This framework allows the customers to build analytical reports, dashboards, and KPIs.
Through ThoughtSpot, The Company provides customers with an out-of-the-box set of reports, dashboards, and KPIs. Also, customers can build reports and dashboards based on their needs and business inquiries and perform advanced AI-driven data insights and searches. Thoughtspot reports and dashboards are integrated with company’s system (the Frontend portal and mobile) applications through the SSO mechanism with white-labeled embedding capabilities to offer the customers and end users one integrated and unified working environment and experience.

\subsection{The DAT model applied to the ADW Case Study}
This section shows the modeling of the ADW case study using DAT. From a structural point of view, Figure \ref{fig:ADW}  shows the primary 4 data nodes; Data Sources, Processing Node (ETL - Keboola), Data warehouse (Snowflake), and Analytics platform (ThoughtSpot).

The Data source shows how to collect the data from different data sources in different formats, integrate with (Time, Finacial, and Sales data), and ingest data to be transferred into a Column-oriented format to be saved on File System(Amazon S3). Other data source is Excel, JSON, and CSV files will be sent directly to the Keboola node to be processed in the cloud.  This Node provides all the needed connectors 
with other data sources. It provides the needed capabilities to schedule (Batch) extractions and transformation. The extracted prepared data is stored in the cloud-based Snowflake data warehouse. Then it will be consumed by another cloud-based advanced BI and analytics framework. The analytics node shows the ability of the framework to provide analytical reports, dashboards, and KPIs to the customers.

\section{Conclusion}
In this work, we presented Data Architecture Tool to model Data Analytical Architecture for Smart Cities Data-Driven Applications  as a part of the VASARI project.
\bibliographystyle{unsrt}  
\bibliography{bib}

 \end{document}